\documentclass[11pt]{scrartcl}
\usepackage{epsfig,amsmath,cite,longtable}
\typearea{9}
\begin{document}

\title{Two electron dynamics in photodetachment}

\author{D. Hanstorp, G. Haeffler, A. E. Klinkm{\"u}ller, U. Ljungblad\\
\textit{Department of Physics, G{\"o}teborg University and}\\
\textit{Chalmers University of Technology AB}\\ 
\textit{SE-412 96 G{\"o}teborg, Sweden}\\
U. Berzinsh\\\textit{Department of Spectroscopy, University of Latvia,}\\
\textit{LV 1586, Riga, Latvia}\\ 
I. Yu. Kiyan\\\textit{Russian Academy of
Sciences, General Physics Institute,}\\ \textit{117 942 Moscow, Russia}\\
D. J. Pegg\\\textit{Department of Physics, University of Tennessee,
Knoxville,}\\ \textit{TN 37996 USA}}

\date{\today}

\maketitle
\begin{abstract}
We present the results of experimental studies of photon-negative ion
interactions involving the dynamics of two electrons. Resonances
associated with doubly excited states of Li$^{-}$ and He$^{-}$ have been
observed using laser photodetachment spectroscopy. Total and partial
photodetachment cross sections have been investigated. In the former
case, the residual atoms are detected irrespective of their excitation
state, while in the latter case only those atoms in specific states are
detected. This was achieved by the use of a state selective detection
scheme based on the resonant ionization of the residual atoms. In
addition, in the case of Li$^{-}$ photodetachment, the threshold
behavior of the Li(2\,$^{2}$P)+e$^{-}$(ks) partial cross section has
been used to accurately measure the electron affinity of Li.
\end{abstract}

\section{Introduction}

We present results of recent studies of the photodetachment of
few-electron atomic negative ions. The motivation for these measurements
is to better understand the general problem of correlated motion of
particles in many-body systems. Correlated motion plays an important
role in several fields of physics. In nuclear physics, for example, it
manifests itself in halo nuclei, in which the two outermost neutrons
contribute to the formation of very diffuse nuclei \cite{one}. In the
case of atomic systems, it is known that negative ions, in particular,
exhibit an enhanced sensitivity to electron correlation effects due to
the suppression of the normally dominant Coulomb force of the core. This
sensitivity is further increased in the case of doubly excited states of
negative ions where, in general, the outermost electron moves in the
field of an excited atom. Such an electron is bound by short range
polarization forces, in contrast to the long range Coulomb force
characteristic for the binding of atoms and positive ions. The more
symmetric the two electron excitation, the more highly correlated the
state. A highly excited pair of electrons are no longer independent,
rather they begin to move collectively. The autodetaching decay of
doubly excited states is manifested as resonances in photodetachment
cross sections. Such resonances arise from an interference between two
paths to the same final continuum state: direct photodetachment and
resonant photodetachment via the intermediate doubly excited state. The
results of experiments on negative ions, and particularly those
involving resonances associated with excited negative ions, therefore
provide sensitive tests of the ability of theory to go beyond the
independent electron model.

The prototypical atomic negative ion is H$^{-}$. Here, two electrons
move in the field of a point core, the proton. In general, if both
electrons are excited the outer electron moves in the field of an
excited H atom consisting of the nucleus and the inner electron. Since
this permanent dipole field differs intrinsically from the long range
Coulomb field, the series of resonances will not be strictly Rydberg in
nature.  Electron correlation becomes more enhanced as the level of
excitation increases since the excited H atom becomes more easily
polarizable. Resonances representing the decay of such doubly excited
states appear in the photodetachment cross section below the excited H
atom thresholds. Double excitation of the H$^{-}$ ion has been studied
extensively both experimentally \cite{two} and theoretically
\cite{three}. It is clear that these highly correlated states are not
well represented by the independent electron model.  Instead, new
quantum number labels have been introduced that describe the collective
motion of the electron pair \cite{four}.  We have chosen to study the
two electron dynamics in He$^{-}$ and Li$^{-}$ ions since they represent
the logical extension of the work on the H$^{-}$ ion. In the case of
both He$^{-}$ and Li$^{-}$, the two active electrons move in the field
of a finite core. He$^{-}$ and Li$^{-}$ ions are therefore simple
systems for investigating few-particle interactions since the core, in
each case, is finite but small. The spherically symmetric closed core of
Li$^{-}$ might be expected to approximate the H$^{-}$ ion somewhat
better than the open core of He$^{-}$. The presence of a core will serve
to lift the degeneracy of the excited state thresholds characteristic of
the H atom. Doubly excited states will then be bound in the shorter
range field of an induced dipole and not a permanent dipole, as is the
case for H$^{-}$. As a consequence, one expects the photodetachment
spectra of these ions to differ from that of the H$^{-}$ ion, at least
at low-to-intermediate levels of excitation. For example, the resonances
should be broader since the lifting of the threshold degeneracy results
in a larger number of continua available for autodetachment
\cite{five,six}. It is also not clear whether the quantum numbers and
propensity rules developed to describe the H$^{-}$ spectra will remain
valid to the same extent in the case of the He$^{-}$ and Li$^{-}$
spectra. Of course, as the level of excitation increases the role played
by the core will become less significant and the spectra, particularly
that of Li$^{-}$, should appear more and more like that of H$^{-}$. In
fact, the Li$^{-}$ ion can be considered as the ``poor man's H$^{-}$
ion'' in the sense that at high levels of excitation their spectra
should be similar. The doubly excited states of Li$^{-}$ are, however,
far more accessible. In both Li$^{-}$ and He$^{-}$, high lying states
can be studied in high resolution experiments using tunable dye
lasers. Corresponding states in the H$^{-}$ ion, however, have
excitation energies about an order of magnitude larger and therefore
remain relatively inaccessible.

\section{Method}

\subsection{Total Cross Sections}

\begin{figure}
\begin{minipage}{\textwidth}
\parbox[b]{0.4\textwidth}{
\epsfig{file=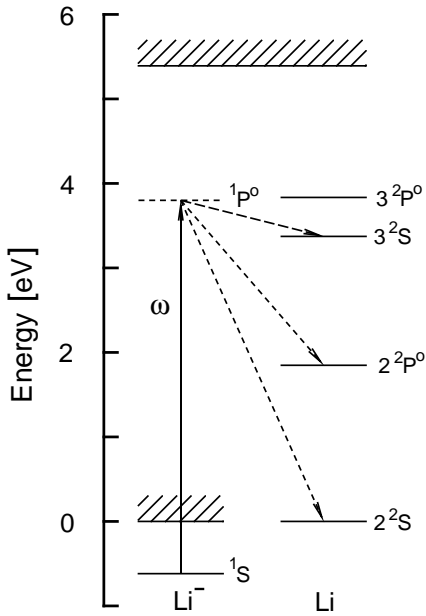, width=0.4\textwidth}
} \hfill
\parbox[b]{0.59\textwidth}{ 
\protect\caption{\label{han-fig01}\sloppy
Partial energy level diagram for the Li$^{-}$/Li systems showing the
autodetaching decay of a doubly excited state of Li$^{-}$ into different
continua.  Each of the three continuum channel is characterized by the
state that the residual Li atom is left in following photodetachment via
the doubly excited state. A particular decay channel is isolated by
state selectively detecting the excited residual atom. } }
\end{minipage}
\end{figure}

The total photodetachment cross section describes the probability that
an electron is detached from a negative ion following the absorption of
a photon, regardless the excitation state of the residual atom or the
energy or direction of the emitted electron.  A total cross section is
the sum of partial cross sections for detachment into each of the
energetically allowed continua. This is illustrated in Fig.\ %
\ref{han-fig01}. Here we show the three possible channels accessible to
a doubly excited state of Li$^{-}$ that lies just below the
Li(3\,$^{2}$P) detachment threshold. The total cross section may be
determined by monitoring the production of either the detached electrons
or the neutral atoms left following photodetachment. In ion beam
experiments the residual atoms can be efficiently collected in the
forward direction since they are moving with the same velocity as the
beam ions. In the present experiments we measure the yield of the
residual atoms produced when a beam of negative ions is merged
collinearly with a beam of pulsed, tunable laser radiation. Studies of
double excitation of negative ions demand both high sensitivity and
energy resolution as a consequence of the low production cross sections
and the high density of resonances below highly excited thresholds. The
choice of a collinear geometry has two major advantages over a crossed
beam geometry. First, the interaction volume can be made at least two
orders of magnitude larger, resulting in a more efficient production of
the doubly excited states. Second, kinematic compression \cite{seven} of
the longitudinal velocity distribution of the ions significantly
improves the energy resolution of the photodetachment spectra.

Typically, doubly excited states autodetach predominantly to just a few
continua and the associated resonance structure modulates the
corresponding partial cross sections. A problem arises in total cross
section measurements as high lying doubly excited states are
accessed. Resonance structure in total cross sections tends to ``wash
out'' as many partial cross sections with little or no structure are
added to those that exhibit well modulated structure. Under such
conditions, it becomes necessary to isolate a particular continuum
channel and study resonances in the corresponding partial cross
section. Another problem arises if one attempts to study the total cross
section for the photodetachment of metastable negative ions such as
He$^{-}$. In this case a large neutral atom background arises from
autodetachment of the unstable ions in their ground state. To avoid
these two problems, we have chosen to selectively ionize the residual
atoms and detect positive ions in the presence of a much smaller
background. By use of this state selective detection scheme, we hence
study partial cross sections instead of total cross sections.

\subsection{Partial Cross Sections}

A partial photodetachment cross section describes the probability that
an electron is detached from a negative ion into a particular continuum
state. Each decay channel is characterized by the excitation state of
the residual atom and the energy and angular momentum of the detached
electron. A particular channel can therefore be identified by either
measuring the energy of the detached electron, as in photoelectron
spectroscopy, or by state selectively detecting the residual atom. We
have chosen the latter alternative in the present work. The excitation
state of the residual atom can be determined by the use of the well
established technique of Resonance Ionization Spectroscopy (RIS). This
sensitive and selective technique was developed more than 20 years ago
\cite{eight} but it has only recently been applied to the study of
negative ions \cite{nine}. The use of a collinear beam geometry combined
with the resonant ionization of the residual atoms presents several
distinct advantages over photoelectron spectroscopy in which the
detached electrons are energy analyzed, invariably using a crossed beam
geometry. First, the positive ions can be collected far more efficiently
than photoelectrons since they move in one direction only, the ion beam
direction. In contrast, the electrons are ejected over a distribution of
angles and therefore only a small fraction can be collected within the
finite acceptance angle of a detector. Second, the background associated
with the detection of positive ions is generally far smaller than that
associated with the detection of photoelectrons.  Both of these
advantages enhance the sensitivity of state selective measurements on
the residual atoms. On the other hand, the major advantage of
photoelectron spectroscopy is that all energetically allowed
photodetachment channels can be isolated and measured
simultaneously. Another attribute is that the angular distribution of
the detached electrons can be measured. This additional information
provides further insight into the photodetachment process.  In order to
extract information on the parameters of a resonance one must fit a
parameter function that describes the interaction of a doubly excited
state with continua to the experimental data. The most commonly used
form is the Beutler- Fano profile \cite{ten}
\begin{align}
\sigma (E) &= a + \frac{(q+\epsilon )^{2}}{1+\epsilon^{2}}\quad
,\label{eq1}\\
\intertext{with,}
\epsilon &= 2\frac{(E-E_{0})}{\Gamma}\quad ,\nonumber
\end{align}
where $E$ is the photon energy, $E_{0}$ and $\Gamma$ are the position
and width of the resonance, $q$ is the shape parameter, $a$ is the
background and $b$ is the amplitude of the resonance. This form is
strictly only valid for the treatment of resonance structure in a total
cross section. It can, however, be used to parametrize resonances in a
partial cross section if, as in the present case, only the positions and
widths of the resonances are required. Starace has extended the
Beutler-Fano treatment to cover the case of partial cross sections
\cite{eleven}. The form that Starace has proposed can be written as
\begin{multline}\label{eq2}
\sigma (E) = a + \frac{b}{1+\epsilon^{2}} \big[ \epsilon^{2} + 2
\epsilon \big( q\Re (\alpha) - \Im (\alpha)\big) + 1 - 2q\Im (\alpha)\\
{}-2\Re (\alpha) + (q^{2}+1)|\alpha |^{2}\big]
\end{multline}
where the parameters have the same significance as in the Beutler-Fano
formula. In addition a complex parameter, $\alpha$, is introduced to
describe the branching into a particular channel.

\subsection{Experimental Procedure}

\begin{figure}
\begin{center}
\epsfig{file=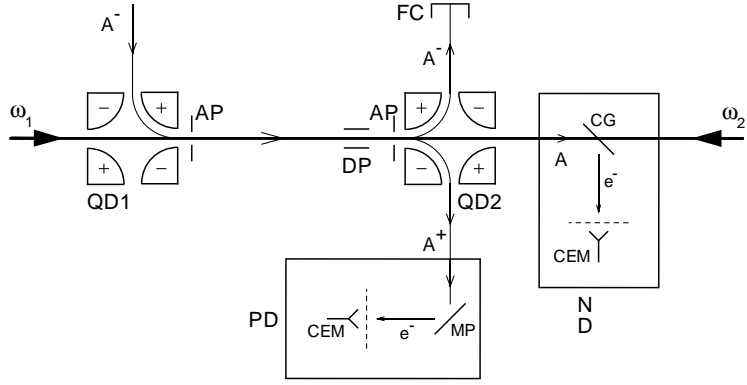, width=0.7\textwidth}
\end{center}
\protect\caption{\label{han-fig02}\sloppy Portion of the collinear
laser-ion beam apparatus. \textsf{QD1,2}, electrostatic quadrupole
deflectors; \textsf{CEM}, channel electron multiplier; \textsf{DP},
deflection plates; \textsf{PD}, positive ion detector; \textsf{FC},
Faraday cup; \textsf{ND}, neutral particle detector; \textsf{CG},
conducting glass plate; \textsf{AP}, aperture; \textsf{MP}, metal plate;
\textsf{A}, any element. The distance between \textsf{QD1} and
\textsf{QD2} is approximately 0.5 m. }
\end{figure}

A detailed description of the interacting beams apparatus used in the
present work can be found elsewhere \cite{twelve}. In short, positive
ions are extracted from a plasma- type ion source and accelerated to
beam energies that can be varied between 2.5~keV and 5~keV. Negative
ions are produced in the beam by double sequential charge exchange in a
cesium vapor. A pair of electrostatic quadrupole deflectors
(\textsf{QD1,2}) is used to direct the negative ion beam into and out of
the path of the laser beams, as shown schematically in Fig.\
\ref{han-fig02}. The ion-laser interaction region is defined by two
apertures (\textsf{AP}) of diameter 3 mm placed 0.5 m apart. The amount
of negative ions remaining after passage through the interaction region
is monitored for normalization purposes using a Faraday cup
(\textsf{FC}) placed adjacent to the second quadrupole
deflector. Neutral atoms produced by photodetachment or collisional
detachment are unaffected by the electric field in the second
quadrupole. They enter the neutral atom detector (\textsf{ND}) situated
downstream of the second quadrupole deflector. Here, the impact of the
fast beam atoms on a quartz plate (\textsf{CG}) produces secondary
electrons that are efficiently detected using a channel electron
multiplier (\textsf{CEM}). The plate is coated with a few atomic layers
of platinum in order to prevent charge build-up. The transparency of the
plate makes it possible to direct the laser beams both parallel and
anti-parallel with respect to the ion beam. By performing both co- and
counter-propagating experiments, it is possible to correct for the
Doppler shift to all orders. A serious background source with this
arrangement, however, is associated with photoelectrons generated on the
plate by pulsed UV radiation. This contribution is greatly reduced by
modulating the bias on a grid placed between the glass plate and the
\textsf{CEM}, as described in detail in \cite{thirteen}. This detector
has been shown to work efficiently for laser wavelengths down to 250 nm.

The third detector (\textsf{PD}) in the interaction chamber is used to
detect positive ions in the state selective measurements. It is situated
adjacent to the second quadrupole, opposite the Faraday cup used to
monitor the negative ion current. It operates in the same way as the
neutral atom detector just described. In this case, however, the
detector is out of the line of sight of the laser beam and so
laser-generated photoelectrons pose no problem. The positive ion
detector is used to state selectively detect residual atoms in the
partial cross section measurements. The electric field of the second
quadrupole is used both to field ionize the Rydberg atoms and to bend
the resulting positive ions into the detector. The positive ion
background is usually small, the major contribution arising from double
detachment collisions of negative ions with the residual gas in the
interaction chamber. This contribution is minimized by a combination of
a good vacuum (approximately 10$^{-9}$~mbar) and a pair of deflection
plates (\textsf{DP}) positioned just prior to the entrance of the second
quadrupole.  The transverse electric field is insufficient to field
ionize the atoms prepared in the highly excited Rydberg states but it
does serve to sweep out of the beam any positive ions produced upstream
in collisional events. This procedure substantially reduces the
collisional detachment background since only atoms left in high lying
Rydberg states will contribute to the background. Of course, this
simultaneously hinders the negative ions from reaching the Faraday
cup. The deflection plates are therefore switched off periodically to
monitor the negative beam intensity.

The laser system used in the experiments consist of two excimer-pumped
dye lasers. The dye lasers can produce narrow bandwidth laser radiation
of wavelengths between 340~nm and 850~nm, which can be extended, with
the use of second harmonic generation, down to 205~nm. The bandwidth is
nominally about 5~GHz, but this can be reduced to 1~GHz by use of an
intra cavity etalon. The energy resolution in the present experiments is
limited solely by the linewidth of the laser. The maximum pulse energy
is typically a few mJ in the visible wavelength region and a few hundred
$\mu$J in the UV, after doubling. The laser intensity is measured after
the passage through the interaction chamber for normalization
purposes. The wavelength of the laser light is calibrated by observing
well known resonance lines in Ne or Ar using optogalvanic spectroscopy
in hollow cathode lamps.

\section{Results}
\subsection{Total Cross Section Measurements}

\begin{figure}\centering
\epsfig{file=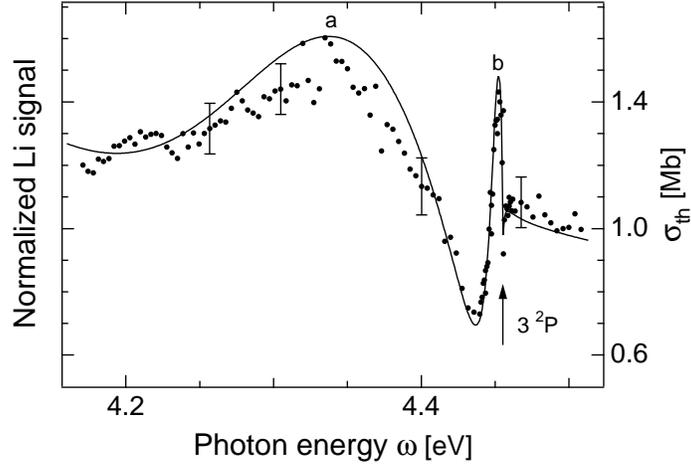, width=0.7\textwidth}
\protect\caption{\label{han-fig03}\sloppy Yield of Li atoms vs. photon
energy in the range 4.2-4.5~eV. The Li signal is proportional to the
total cross section for the photodetachment of Li$^{-}$ via the
2\,$^{2}$Skp, 2\,$^{2}$Pks,d and 3\,$^{2}$Skp channels (see Fig.\ %
\ref{han-fig01}). The experimental data (dots) has been normalized to
theory (solid line with scale to the right) at the 3\,$^{2}$P threshold. }
\end{figure}

The cross section for photodetachment of an electron from the Li$^{-}$
ion has previously been investigated, both experimentally
\cite{fourteen} and theoretically \cite{fifteen}. Earlier laser studies
have focused on the region around the first and second detachment
thresholds. The latter, also called the first excited state threshold,
corresponds to the process in which the residual Li atom is left in the
2\,$^{2}$P state following photodetachment. In contrast to the heavier
alkali ions \cite{sixteen}, the Li$^{-}$ photodetachment cross section
shows no resonance structure below the first excited state threshold. It
is expected, however, that doubly excited states should appear at higher
levels of excitation. The atomic wave function then becomes more
extended in space, making it easier for an extra electron to share the
attractive force from the nucleus, or equivalently, the atom becomes
more polarizable. We therefore initiated a search for doubly excited
$^{1}$P$^{\text{o}}$ states of Li$^{-}$, that are bound with respect to
the 3\,$^{2}$P state of the Li atom \cite{seventeen}. The results in the
energy region 4.2-4.5~eV are shown in Fig.\ \ref{han-fig03}. The Li atom
signal is proportional to the total cross section which, in this case,
is equal to the sum of the partial cross sections for the three channels
shown in Fig.\ \ref{han-fig01}. The experimental cross section data are
not absolute. The measurements have therefore been normalized to a
theoretical curve at the Li(3\,$^{2}$P) threshold. The main contribution
to the approximately 10\% scatter in the experimental data arises from
changes in the spatial overlap of the laser and ion beams that occurred
as the laser frequency was scanned. The statistical scatter due to
counting of the residual atoms was less than 3\%. The experimental data
exhibits three significant features. A small narrow dip is seen at
4.456~eV, a rather narrow structure is observed at 4.453~eV (labeled b)
and finally, a broad resonance is seen to cover essentially the entire
energy region between the Li(3\,$^{2}$S) and Li(3\,$^{2}$P) thresholds
(labeled a).

The small dip at 4.456~eV is a Wigner cusp that arises from the opening
of the Li(3\,$^{2}$P)+e$^{-}$(ks) photodetachment channel (hereafter
labeled 3\,$^{2}$Pks, for brevity). The other two structures are
resonances associated with the decay of doubly excited states. These
resonances overlap one another as well as the Wigner cusp, making it
impossible to extract information on individual positions and
widths. The Beutler-Fano formula, which is normally used to extract
resonance parameters, cannot be applied in this case since it is only
valid for isolated resonances. In order to explain the origin of the
observed structures in the Li$^{-}$ photodetachment cross section,
comparisons with two independent calculated cross sections were
made. Lindroth \cite{seventeen} used an approach which combines complex
rotation \cite{eighteen} with a discrete numerical basis set
\cite{nineteen}, whereas Pan et al. \cite{five} employed the R-matrix
method.  The two theoretical methods yield essentially identical results
(see the solid line in Fig.\ \ref{han-fig03}). We observe a rather good
agreement between theory and experiment over the whole energy region
studied. According to Lindroth«s calculation, the broad structure (a) is
identified with an intrashell resonance associated with the symmetrical
excitation of the two valence electrons. This resonance was found to be
dominated by the 3p3d and 4s3p configurations. The narrow structure (b)
at approximately 4.453~eV corresponds to several overlapping intershell
resonances in which the valence electrons are asymmetrically excited.

\subsection{Partial Cross Section Measurements}

The state selective detection scheme based on RIS has been applied to
isolate specific autodetaching channels for doubly excited states in
both He$^{-}$ and Li$^{-}$. Here we present the results of recent
measurements of the positions and widths of doubly excited states which
appear as resonances in partial cross sections below certain excited
state thresholds. The 1s3s4s\,$^{4}$S state in He$^{-}$, for example,
lies just below the He(3\,$^{3}$S) threshold. A Feshbach resonance
associated with the autodetaching decay of this state was observed in
both the He(1s2s\,$^{3}$S)+e(ks) and He(1s2p\,$^{3}$P)+e(kp) partial
cross sections (hereafter labeled 2\,$^{3}$Sks and 2\,$^{3}$Pkp)
\cite{twenty}. Similarly, in the case of Li$^{-}$, resonances due to
doubly excited states below the Li(4\,$^{2}$P) and Li(5\,$^{2}$P)
thresholds were observed in the 3\,$^{2}$Skp partial cross section
\cite{twentyone}. The state selective detection scheme has also been
used to study the threshold behavior of the 2\,$^{2}$Pks partial cross
section. This technique enabled Haeffler et al.\ \cite{twentytwo} to
accurately measure the electron affinity of Li.

\subsubsection{Doubly Excited States of He$^{-}$}

Doubly excited states of He$^{-}$ of doublet symmetry have been observed
in studies of electron impact on He \cite{twentythree}. In contrast,
data on quartet states are sparse. Selection rules on photoexcitation
from the $^{4}$P$^{\text{o}}$ ground state limit excited state
production to those of $^{4}$S, $^{4}$P and $^{4}$D symmetry. Recently,
the He$^{-}$ photodetachment cross section below 4~eV has been
calculated by Xi and Froese Fischer \cite{twentyfour}. They predicted a
number of resonances associated with the decay of doubly excited quartet
states.  Bae and Peterson \cite{twentyfive} observed a
(1s2p$^{2}$\,$^{4}$P) shape resonance in the total photodetachment cross
section of He$^{-}$ just above the He(2\,$^{3}$P) threshold. The
parameters of this resonance were subsequently measured more accurately
by Walter et al. \cite{twentysix}. We recently embarked on a program
aiming to investigate doubly excited states in He$^{-}$ below the
He($n>2$) thresholds. The first measurement involved the 1s3s4s\,$^{4}$S
state that lies below the He(3\,$^{3}$S) threshold. The decay of this
state was observed as a Feshbach resonance in both the 2\,$^{3}$Sks and
the 2\,$^{3}$Pkp partial cross sections.

\begin{figure}
\begin{minipage}{\textwidth}
\parbox[b]{0.48\textwidth}{
\epsfig{file=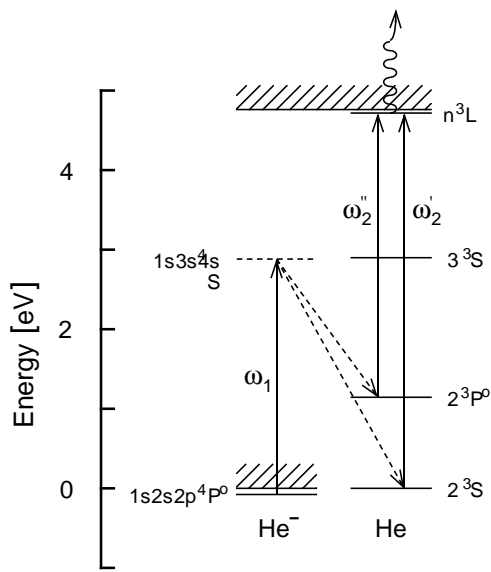, width=0.46\textwidth}
} \hfill
\parbox[b]{0.51\textwidth}{ \protect\caption{\label{han-fig04}\sloppy
Partial energy level diagram for the He$^{-}$/He systems. Solid lines
represent photoexcitation, the wavy line field ionization and dashed
lines autodetachment. The 1s3s4s\,$^{4}$S state is produced by
absorption of photons of frequency $\omega_{1}$. The 2\,$^{3}$Sks and
2\,$^{3}$Pkp decay channels were isolated by selectively detecting the
residual He(2\,$^{3}$S) and He(2\,$^{3}$P) atoms using RIS (laser
$\omega_{2}^{\prime}$  and $\omega_{2}^{\prime\prime}$,
respectively). } }
\end{minipage}
\end{figure}

Fig.\ \ref{han-fig04} shows selected energy levels of the He$^{-}$/He
systems. The 1s3s4s\,$^{4}$S state of He$^{-}$ is situated just below
the He (3\,$^{3}$S) threshold. This state, which is excited with laser
$\omega_{1}$, rapidly autodetaches via the 2\,$^{3}$Sks and 2\,$^{2}$Pkp
channels. Following the decay, the residual He atom will be left in
either the 2\,$^{3}$S or 2\,$^{3}$P excited states.  Two different laser
frequencies $\omega_{2}^{\prime}$ and $\omega_{2}^{\prime\prime}$ were
applied separately in the resonance ionization scheme used to monitor
the population of the 2\,$^{3}$S and 2\,$^{3}$P states. The frequency
was chosen to induce a transition between the 2\,$^{3}$S and the
24\,$^{3}$P states of the He atom, when photodetachment into the
2\,$^{3}$Sks channel was studied. The frequency
$\omega_{2}^{\prime\prime}$ induced a resonance transition between the
2\,$^{3}$P and 26\,$^{3}$D states of He, when photodetachment into the
2\,$^{3}$Pkp channel was studied. The population of both the high lying
Rydberg states were efficiently depleted by the electric field of the
second quadrupole deflector and He$^{+}$ ions thus produced were
recorded as a function of frequency of laser $\omega_{1}$. The output of
laser $\omega_{1}$ was attenuated to avoid saturation of the
photodetachment process and the detector.  

\begin{figure}\centering
\epsfig{file=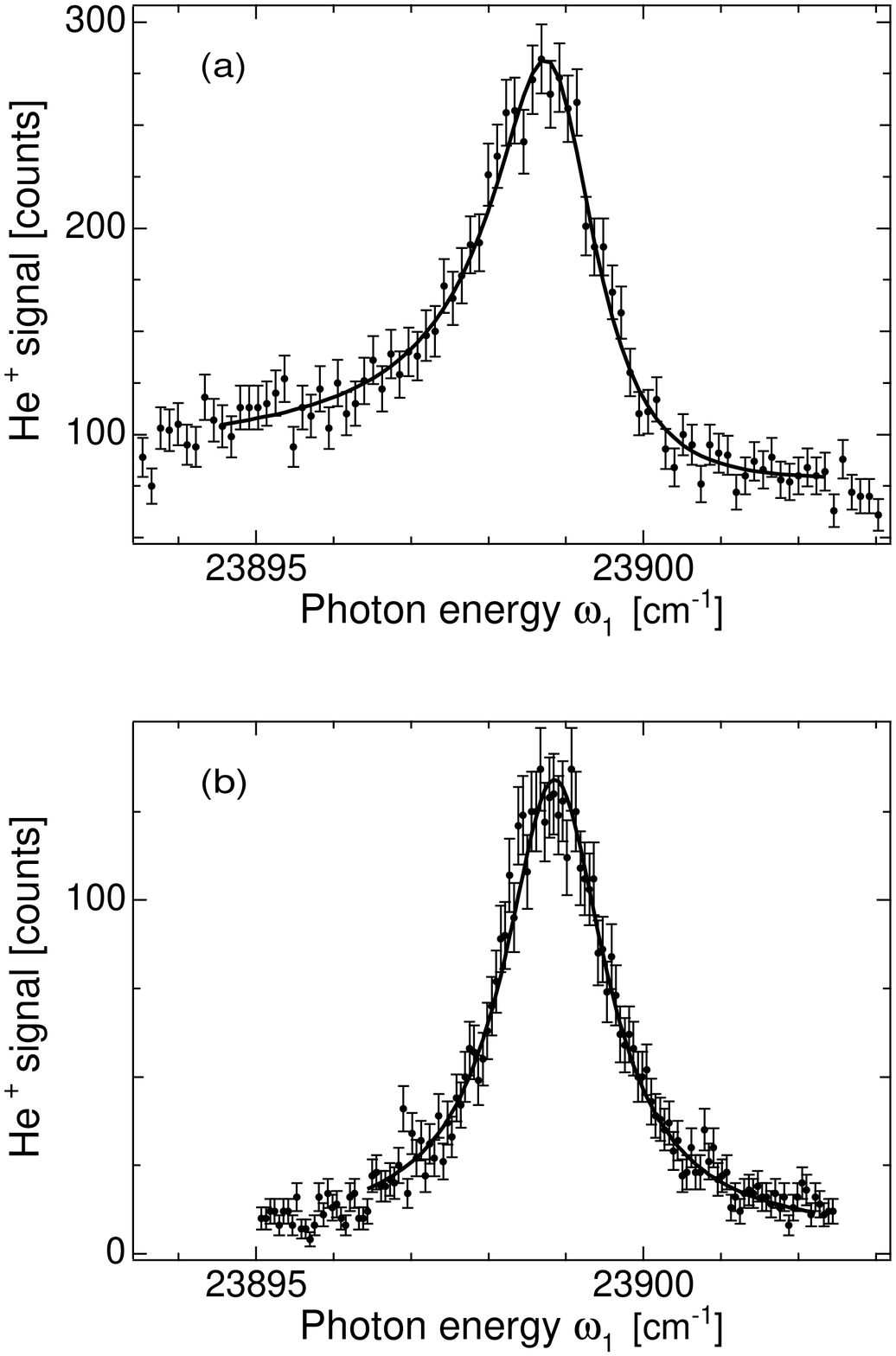, height=0.8\textheight}
\protect\caption{\label{han-fig05}\sloppy
Yield of He$^{+}$ ions vs. photon energy. The He$^{+}$ signal is
proportional to the partial cross section for the photodetachment of
He$^{-}$ via the 2\,$^{3}$Sks,d channel (a) and the 2\,$^{3}$Pkp channel
(b). The solid line represents a fit of the Fano function \eqref{eq1} to
the experimental data (dots).}
\end{figure}

Fig.\ \ref{han-fig05} shows how the He$^{+}$ signal varies as a function
of the laser frequency $\omega_{1}$ in the vicinity of the
resonance. The signal is proportional to the partial cross sections for
photodetachment via the 2\,$^{3}$Sks,d channels (Fig.~\ref{han-fig05}a)
and the 2\,$^{3}$Pkp channel (Fig.~\ref{han-fig05}b). Selection rules
forbid the autodetachment of the 1s3s4s\,$^{4}$S state via the
2\,$^{3}$Skd channel in the LS-coupling approximation and so in Fig.
\ref{han-fig05} (a) this channel contributes to the background only (we
do not resolve the degenerate s- and d-wave channels). According to
recent calculations of Xi and Froese Fischer \cite{twentyseven} the
cross section for the d-wave channel remains nearly constant at 6~Mb
across the resonance, while the s-wave channel has a peak value of 20~Mb
on the resonance.  The difference in the level of the background in
Fig. \ref{han-fig05} (a) and Fig. \ref{han-fig05} (b) arises essentially
from the aforementioned d-wave channel. The background in
Fig. \ref{han-fig05} (b), and that remaining in Fig. \ref{han-fig05} (a)
after the d-wave contribution has been subtracted, arises from several
sources. The two most significant background processes that produce
He$^{+}$ ions, indirectly from He$^{-}$ ions, are collisional single
detachment and photodetachment processes by laser $\omega_{2}^{\prime}$
($\omega_{2}^{\prime\prime}$) that populate the He(2\,$^{3}$S,
2\,$^{3}$P) states.  Once these states are formed by background
processes, they will be resonantly ionized by the laser frequencies
$\omega_{2}^{\prime}$ and $\omega_{2}^{\prime\prime}$, respectively, and
the resulting He$^{+}$ ions will be indistinguishable from the
signal. The He$^{+}$ ions produced directly in double detachment events
were essentially eliminated in our experiment by the use of a pair of
deflection plates placed just upstream of the second quadrupole
deflector (see Fig. \ref{han-fig02}). The low pressure in the
interaction chamber is essential to reduce the background due to
collisional detachment processes. In addition, the output of the laser
at frequencies $\omega_{2}^{\prime}$ and $\omega_{2}^{\prime\prime}$ was
attenuated to reduce the background arising from the photodetachment of
He$^{-}$ at these frequencies.  

\setlongtables
\begin{longtable}[c]{lr@{.}lr@{.}ll}
\caption{\label{ta1}\slshape A comparison of experimental and
theoretical values for the positions and widths of resonant states in
He$^{-}$  and Li$^{-}$ relative to the ground state of the negative
ion.}\\
\hline\hline
Resonance & \multicolumn{2}{c}{$E_{0}$ (eV)} &
\multicolumn{2}{c}{$\Gamma$ (eV)} & Reference\\ \hline
\endfirsthead
\multicolumn{6}{r}{continued from previous page}\\
\hline\hline
Resonance & \multicolumn{2}{c}{$E_{0}$ (eV)} &
\multicolumn{2}{c}{$\Gamma$ (eV)} & Reference\\ \hline
\endhead
\hline\hline
\multicolumn{6}{r}{continued on next page}
\endfoot
\hline\hline
\endlastfoot
He$^{-}$($^{4}$S) & 2&959\,255(7) & 0&000\,19(3) & Experiment
\cite{twenty} \\
 & 2&959\,07 & 0&000\,19 & Theory \cite{twentyfour} \\[0.5ex]
Li$^{-}$(c) & 5&113\,2(4) & 0&007\,4(5) & Experiment \cite{twentyone}\\ 
 & 5&115 & 0&007\,3 & Theory \cite{six} \\[0.5ex]
Li$^{-}$(d) & 5&123\,4(4) & 0&007\,6(11) & Experiment \cite{twentyone}
\\ 
 & 5&126 & 0&009\,9 & Theory \cite{six} \\[0.5ex]
Li$^{-}$(j) & 5&448\,5(1) & 0&001\,8(1) & Experiment \cite{twentyone} 
\\[0.5ex]
Li$^{-}$(k) & 5&450\,8(1) & 0&007\,3(3) & Experiment \cite{twentyone}
\end{longtable}

The resonance energy and width are obtained by fitting the data shown in
Fig.~\ref{han-fig05} to both a Beutler-Fano profile \eqref{eq1} and to
the form proposed by Starace \eqref{eq2}.  As expected, the values of
$E_{0}$ and $\Gamma$ were the same using both parametrizations. In each
case, the measurement was performed with co-propagating and counter-
propagating laser ($\omega_{1}$) and ion beams to eliminate the Doppler
effect associated with the fast moving ions. The data shown in
Fig. \ref{han-fig05}, for instance, have been recorded with
co-propagating laser ($\omega_{1}$) and ion beams. The fit to these data
yields the blue-shifted resonance position, $E_{0}^{\text{b}}$. The
measurement was repeated with counter- propagating beams to yield the
red-shifted resonance position, $E_{0}^{\text{r}}$. The resonance
position, corrected for the Doppler effect to all orders, is given by
the geometric mean, $E_{0}=\sqrt{E_{0}^{\text{b}} E_{0}^{\text{r}}}$, of
the red- and blue-shifted energies. Table \ref{ta1} compares the present
measurement of the resonance position and width with the result of a
recent calculation of Xi and Froese Fischer \cite{twentyfour}. One can
see that the present results are in good agreement with theory.  

\subsubsection{Doubly Excited States in Li-}

\begin{figure}
\begin{minipage}{\textwidth}
\parbox[b]{0.5\textwidth}{
\epsfig{file=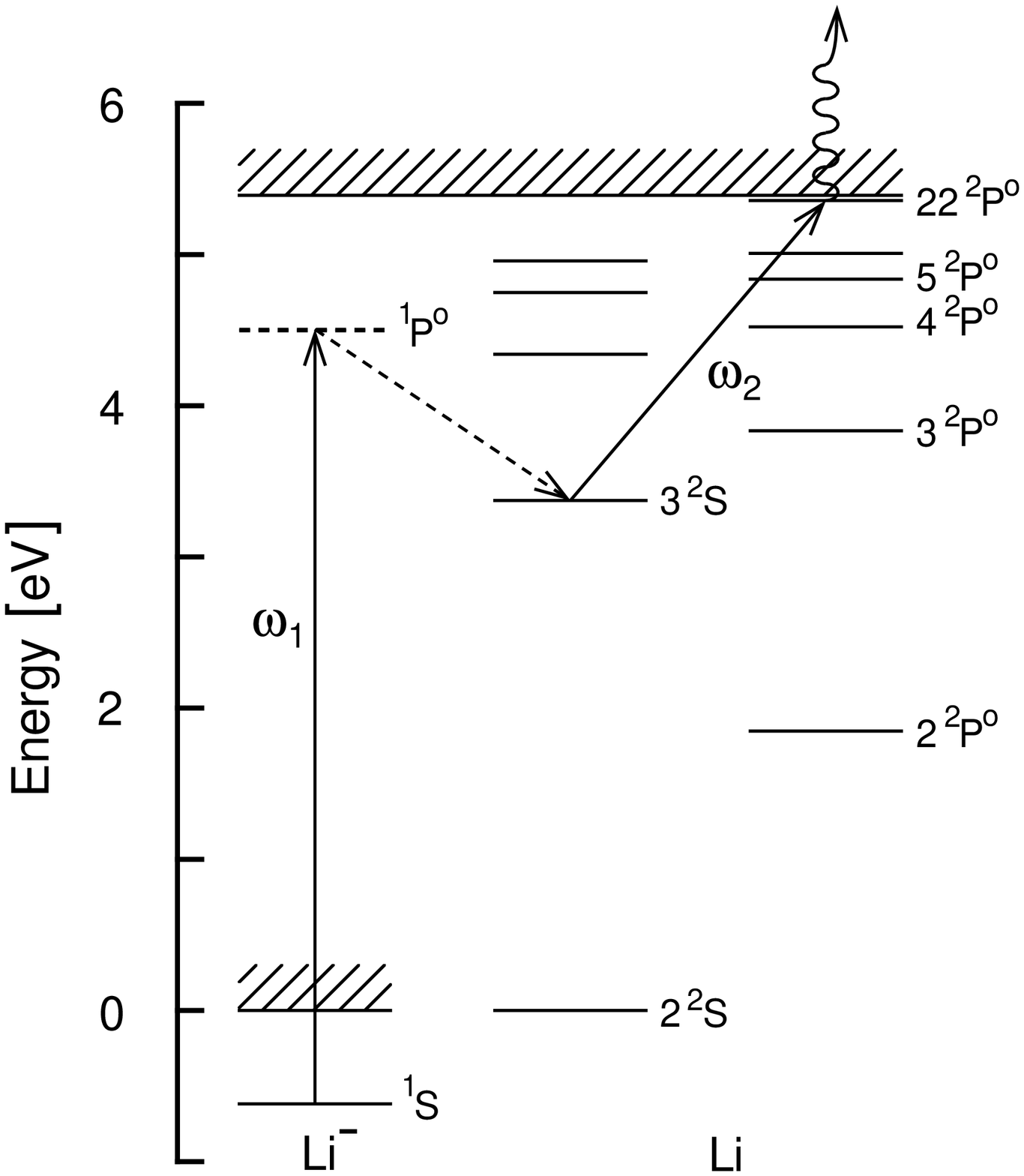, width=0.48\textwidth}
} \hfill
\parbox[b]{0.49\textwidth}{ \protect\caption{\label{han-fig06}\sloppy
Partial energy level diagram for the Li/Li$^{-}$ systems. Solid lines
represent photoexcitation, the wavy line field ionization and the dashed
line autodetachment via the 3\,$^{2}$Skp channel. Doubly excited states
of Li$^{-}$ are produced by the absorption of photons of frequency
$\omega_{1}$. The
3\,$^{2}$Skp partial cross section is isolated by selectively detecting
the residual atoms left in the 3\,$^{2}$S state using RIS (laser
$\omega_{2}$). } }
\end{minipage}
\end{figure}

The method of studying partial photodetachment cross sections based on
the selective detection of the residual atoms has also been used to
investigate resonances associated with the decay of doubly excited
states of Li$^{-}$. Fig. \ref{han-fig06} illustrates a state selective
detection scheme designed, in this case, to isolate the 3\,$^{2}$Skp
partial cross section in Li$^{-}$ photodetachment. A laser of frequency
$\omega_{1}$ excites the negative ion into a $^{1}$P doubly excited
state that is embedded in several continua. This transient state
subsequently autodetaches producing a free electron and a neutral Li
atom which, in general, is excited. In the case shown in the figure, the
quasi-discrete state is embedded in six continua i.e. there are six
possible decay channels that are energetically accessible. These
channels are characterized by the residual Li atom being left in
2\,$^{2}$ S, 2\,$^{2}$P, 3\,$^{2}$S, 3\,$^{2}$P, 3$^{2}$D or 4$^{2}$S
state following detachment. A second laser of frequency $\omega_{2}$ is
applied to resonantly photoexcite those Li atoms left in the 3\,$^{2}$S
state to the 22\,$^{3}$P state. Atoms in such highly excited Rydberg
states are efficiently ionized by the electric field of the second
quadrupole. The positive ions thus produced can be efficiently detected
in the presence of a relatively low background. This procedure isolates
the 3\,$^{2}$Skp channel.

\begin{figure}\centering
\epsfig{file=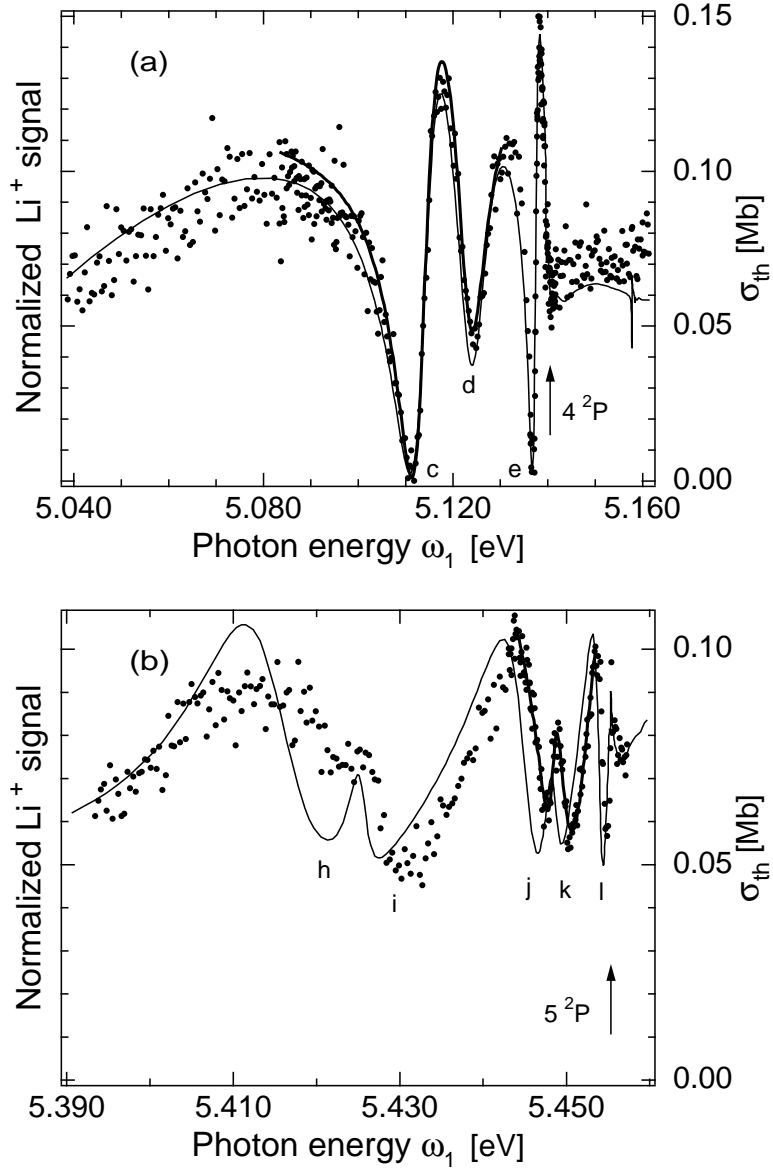, height=0.8\textheight}
\protect\caption{\label{han-fig07}\sloppy Yield of Li$^{+}$ ions
vs. photon energy in the ranges 5.04-5.16~eV (Fig.  \ref{han-fig07}a)
and 5.39-5.46~eV (Fig. \ref{han-fig07}b). The Li$^{+}$ signal is
proportional to the partial cross section for photodetachment of
Li$^{-}$ via the 3\,$^{2}$Skp channel (see Fig. \ref{han-fig06}).  The
experimental data (dots) has been normalized to theory (solid line with
scale at the right). The thick solid lines indicate fits of the double
Fano formula \eqref{eq3} to the data.  }
\end{figure}

In Fig. \ref{han-fig07}, we show the cross section for photodetachment
of Li$^{-}$ via the 3\,$^{2}$Skp channel over photon energies of
approximately 5.04-5.16~eV and 5.29-5.46~eV. These ranges cover the
regions below, and including, the Li(4\,$^{2}$P) and Li(5\,$^{2}$P)
thresholds, respectively. Several Feshbach ``window'' resonances are
observed to lie below these thresholds. In the figure the present
measurements are compared with the result of a recent eigenchannel
R-matrix calculation by Pan et al. \cite{twentyeight}. The experimental
resolution, which is estimated to be about 25~$\mu$eV, is sufficiently
high compared to the typical resonance widths that a direct comparison
with theory can be made without resorting to deconvolution
procedures. At photon energies just below 5.39~eV, the ionization energy
of Li, we also observed in-board calibration lines arising from the
photoexcitation of ground state Li atoms, by laser $\omega_{1}$, to high
lying Rydberg states that are subsequently field ionized. It was
unnecessary in this case to resort to the use of co- and
counter-propagating beams to eliminate the Doppler effect since the
resonances in the Li$^{-}$ spectra are much broader than those in the
He$^{-}$ spectra. The calibration procedures described above sufficed to
determine the resonance position. The cross section measurements are
relative. In Fig. \ref{han-fig07}a and \ref{han-fig07}b, they have been
normalized to the theoretical results of Pan et al. \cite{twentyeight}
by multiplying the data by factors equal to the ratio of the areas under
the experimental and theoretical cross section curves between the limits
5.04-5.16~eV and 5.39-5.46~eV, respectively.

Fig. \ref{han-fig07} shows that there is generally a good qualitative
agreement between experiment and theory in the sense that all the
predicted resonances have been observed and no new resonances are
apparent. There is also excellent quantitative agreement in the case of
the resonant structure below the Li(4\,$^{2}$P) threshold, between
5.04~eV and 5.16~eV (Fig. \ref{han-fig07}a). The situation is slightly
different, however, for the resonances below the Li(5\,$^{2}$P)
threshold. Here, theory fails to predict the exact energies and
strengths of the resonances. For example, there exists a difference in
energy between the measured and calculated resonances ranging from
0-2~meV. The size of the discrepancy appears to increase the further the
resonance is from the threshold. Experiment and theory overlap, however,
close to the threshold. The resonance labeled h is also seen to be
considerably weaker than theoretically predicted, while resonance j is
not as strong as resonance k, contrary to theory. The more recent
calculation by Liu and Starace \cite{twentynine} shows, however, an
improved agreement with the experimental data.

The window resonances shown in Fig. \ref{han-fig07} partially overlap
each other, and it is therefore not possible to fit a single Fano
profile to each resonance. We have instead fitted a form representing
the sum of two Fano profiles to the data in order to extract the energy
and width of each resonance pair:
\begin{align}\label{eq3}
\sigma (E) &= a + \sum_{n=1}^{2} b_{n} \frac{(q_{n} +
\epsilon_{n})^{2}}{1 + \epsilon_{n}^{2}}\quad ,\\
\intertext{with}
\epsilon_{n} &= 2 \frac{(E-E_{0n})}{\Gamma}\quad .
\end{align}
Here, the index $n$ represent the parameters for each resonance. Using
this method we determined the parameters for resonances c and d in
Fig. \ref{han-fig07}a. The measured positions and widths are shown in
Table \ref{ta1}, along with corresponding values calculated by Lindroth
\cite{six}. There is a good agreement between the experiment and theory
in this case. Lindroth's resonance parameters are derived directly from
a complex rotation calculation. The R-matrix calculation of Pan et
al. \cite{twentyeight} did not explicitly yield the resonance parameters
and therefore cannot be used for comparison. Since the Fano formula
strictly only applies to total cross sections, the values of the $q$
shape parameters are not entirely meaningful in the context of partial
cross sections. This parameter is therefore omitted in the table.

We also attempted, unsuccessfully, to include the resonance labeled e in
the fitting procedure. In this case it appears that the resonance is
prematurely terminated by the opening of the 4\,$^{2}$Pks channel. The
resonance energies in this case depended strongly on the interval of the
fit, which was not the case when the c and d resonances were treated as
a pair. In a similar manner, the double Fano formula was fit to the two
resonances labeled j and k. The parameters obtained from this fit are
also included in Table \ref{ta1}. In this case there is no theoretical
data available for comparison.

\begin{figure}\centering
\epsfig{file=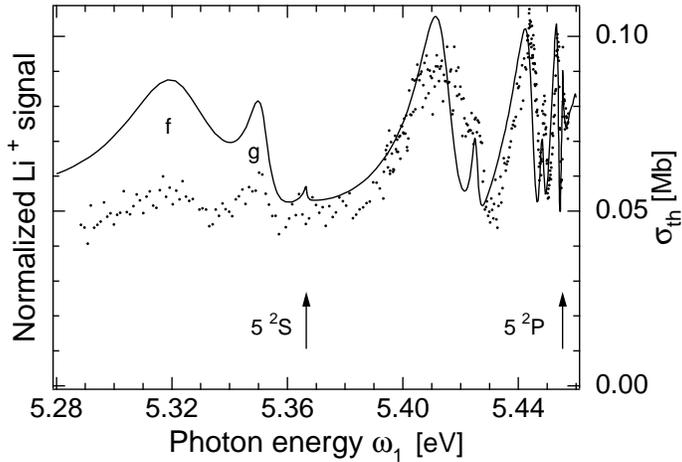, width=0.7\textwidth}
\protect\caption{\label{han-fig08}\sloppy Yield of Li$^{+}$ ions
vs. photon energy in the ranges 5.29-5.46~eV (Fig.
\ref{han-fig07}b). The Li$^{+}$ signal is proportional to the partial
cross section for photodetachment of Li$^{-}$ via the 3\,$^{2}$Skp
channel (see Fig. \ref{han-fig06}). The experimental data (dots) has
been normalized to theory (solid line with scale at the right).  }
\end{figure}

In Fig. \ref{han-fig08} we show an extended 3\,$^{2}$Skp partial
photodetachment cross section which includes measurements below the
Li(5\,$^{2}$S) threshold in addition to the data already shown in
Fig. \ref{han-fig07}b. The data in the range 5.29-5.39~eV represents a
relatively low statistics survey scan. It is normalized to the
calculation of Pan et al.\cite{twentyeight} using the same factor as for
the data between 5.39-5.46~eV. It is clear, even from this relatively
low quality data, that the two resonances labeled f and g are observed
at approximately the calculated energies but their measured strengths
appear to be weaker than predicted. Presumably, the lowest lying
resonance, labeled f, is the intrashell resonance representing symmetric
excitation of the two valence electrons.

\subsubsection{Threshold Investigations of Li$^{-}$}

Wigner \cite{thirty} predicted that in the vicinity of a threshold the
photodetachment cross section is represented by
\begin{equation}\label{eq4}
\sigma (E) \sim \begin{cases} (E-E_{0})^{l+1/2} & E \ge E_{0}\\
				0 & E<E_{0}\quad ,
		\end{cases}
\end{equation}
where $E-E_{0}$ is the excess energy above threshold and $l$ is the
angular momentum of the outgoing electron. This form has been tested
experimentally \cite{thirtyone} and is now well established. The range
of validity of the Wigner law describing threshold behavior, however,
varies from system to system. It is primarily determined by the
polarizability of the residual atom \cite{thirtytwo}. Resonances near
the threshold also play a role in limiting the energy region over which
the threshold law applies. If the outgoing electron is represented by
$l$=0 (s-wave detachment) the cross section rises with an infinite slope
at the threshold. This allows one, in principle, to perform very
accurate threshold measurements. If, however, $l\ge 1$ the slope of the
cross section is zero at threshold, making high precision measurements
much more difficult.

\begin{figure}\centering
\epsfig{file=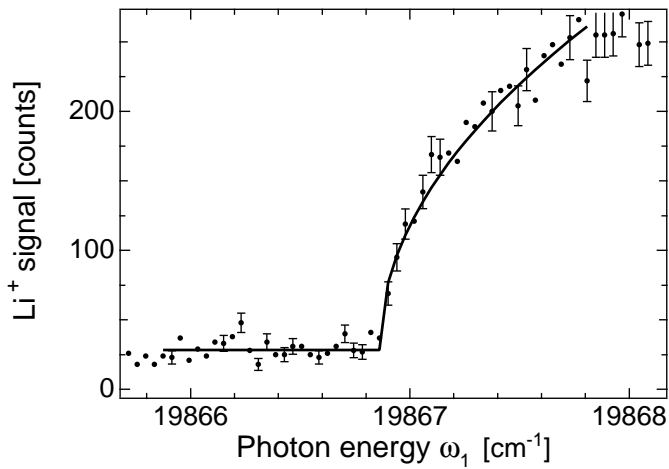, width=0.7\textwidth}
\protect\caption{\label{han-fig09}\sloppy Yield of Li$^{+}$ ions
vs. photon energy in the region of the 2\,$^{2}$P threshold.  The
Li$^{+}$ signal is proportional to the partial cross section for
photodetachement of Li$^{-}$ via the 2\,$^{2}$Pks channel (See
Fig. \ref{han-fig01}). The solid line is a fit of the Wigner law
\eqref{eq4} to the data in the range shown. The error bars on selected
data points represent the shot noise.  }
\end{figure}

An investigation of the threshold behavior of the photodetachment
process in a s-wave detachment channel, using tunable laser sources, has
the potential for yielding the most accurate values for electron
affinities. It was recently demonstrated \cite{twentytwo,thirtythree}
that, by state selectively detecting the residual atom following
photodetachment, it is possible to determine an electron affinity to
very high accuracy for, in principle, almost any element. In such a
scheme, the residual atoms are resonantly ionized and the resulting
positive ions are detected in the presence of a low background. The
excellent signal-to-background ratio in this case produces data of high
statistical quality, which is especially valuable in the vicinity of a
threshold. In addition, the use of a collinear laser-ion beam geometry
results in high energy resolution measurements. Again, this is
particularly valuable near a threshold where the range of validity of
the Wigner law is limited. We demonstrated the potential of this
technique \cite{twentytwo} in a measurement of the electron affinity of
Li. The result is about an order of magnitude more accurate than any
previous measurement. In the experiment we selectively detected the
residual Li atoms left in the 2\,$^{2}$P state following photodetachment
via the 2\,$^{2}$Pks channel. The sharp onset of production of Li$^{+}$
ions at this s-wave detachment threshold enabled us to easily fit the
Wigner law form to the near-threshold data and extrapolate to the
threshold value. The data shown in Fig. \ref{han-fig09} were taken with
counter-propagating laser and ion beams. The measurement was repeated
with co-propagating beams. The threshold energy was determined,
corrected for the Doppler effect to all orders, from the geometrical
mean of the measured blue- and red-shifted threshold energies. The
energy of the threshold of the 2\,$^{2}$Pks channel was found to be
19\,888.55(16)~cm$^{-1}$. The well known separation of the
2\,$^{2}$S-2\,$^{2}$P states of Li \cite{thirtyfour} was subtracted from
the threshold energy to yield an electron affinity of 618.049(20)~meV
using the conversion factor 8065.5410~(cm$^{-1}$/eV).

\section{Conclusions}

The structures of singly excited atoms and positive ions are well
described within the independent particle approximation. A large base of
experimental data is available for such systems. More loosely bound
systems such as negative ions and doubly excited systems in general are,
however, more sensitive to electron correlation since here the Coulomb
field becomes less dominant. In these cases, the independent electron
model begins to break down and collective motion becomes
apparent. Experimental data on doubly excited states of negative ions,
where correlation effects are of paramount importance, are relatively
sparse. Until a few years ago, for example, a systematic investigation
of doubly excited states had only been made in the case of the simplest
two electron negative ion, H$^{-}$.  

In the present work we have extended the study of double excitation to
the He$^{-}$ and Li$^{-}$ ions which also have two active valence
electrons. We have determined the positions and widths of several states
in these systems by analyzing resonances in photodetachment cross
sections. The measurements were performed under conditions of high
sensitivity and resolution since production cross sections are typically
small and resonances are close lying in energy. The sensitivity and
resolution were enhanced by combining a collinear laser-ion beam
apparatus with the detection of positive ions produced by resonantly
ionizing the residual atoms from the photodetachment process. This state
selective detection scheme, which is based on the well established
technique of Resonance Ionization Spectroscopy, greatly extends the
field of negative ion spectroscopy. It is our intention to continue
these studies and investigate doubly excited states in He$^{-}$ and
Li$^{-}$ and their branching into different continua, all the way to the
double detachment limits.

\section{Acknowledgements}

We would like to take the opportunity to acknowledge the support and
encouragement that we have received from Professor Ingvar Lindgren
throughout this work. As a theorist, his interest in our work most
certainly stems from the enhanced role played by electron correlation in
negative ion physics. As a former practicing experimentalist, however,
we suspect he derives great pleasure from observing how the development
of new experimental techniques allows investigations of more and more
subtle effects in atomic systems. Thanks Ingvar, we wish you well in
your future pursuits.  We also thank E. Lindroth, A. F. Starace,
C.-N. Liu, J. Xi and C. F. Fischer for fruitful discussions and for
providing us with unpublished data. Financial support for this research
has been obtained from the Swedish Natural Science Research Council
(NFR). DJP acknowledge support from the Swedish Institute, The Royal
Swedish Academy of Sciences and the US Department of Energy, Office of
Basic Energy Sciences, Division of Chemical Sciences. UB acknowledge
personal support from the Swedish Institute and, finally, IYK
acknowledge support from the Wenner-Gren Center Foundation.


\end{document}